\documentclass[%
article,
superscriptaddress,
nofootinbib,
amsmath,amssymb,
aps,
pra,
notitlepage,
showkeys]{revtex4-2}

\usepackage[margin=3cm]{geometry}
\usepackage{color}
\usepackage{url}
\usepackage{graphicx}
\usepackage{lipsum}
\usepackage[export]{adjustbox}
\usepackage[font=footnotesize,labelfont=bf]{caption}
\usepackage{mathtools}
\usepackage{gensymb}
\usepackage[T1]{fontenc}
\usepackage{amsmath}
\usepackage{amssymb}

\usepackage{scalerel}
\usepackage{tikz}
\usetikzlibrary{svg.path}

\definecolor{orcidlogocol}{HTML}{A6CE39}
\tikzset{
  orcidlogo/.pic={
    \fill[orcidlogocol] svg{M256,128c0,70.7-57.3,128-128,128C57.3,256,0,198.7,0,128C0,57.3,57.3,0,128,0C198.7,0,256,57.3,256,128z};
    \fill[white] svg{M86.3,186.2H70.9V79.1h15.4v48.4V186.2z}
                 svg{M108.9,79.1h41.6c39.6,0,57,28.3,57,53.6c0,27.5-21.5,53.6-56.8,53.6h-41.8V79.1z M124.3,172.4h24.5c34.9,0,42.9-26.5,42.9-39.7c0-21.5-13.7-39.7-43.7-39.7h-23.7V172.4z}
                 svg{M88.7,56.8c0,5.5-4.5,10.1-10.1,10.1c-5.6,0-10.1-4.6-10.1-10.1c0-5.6,4.5-10.1,10.1-10.1C84.2,46.7,88.7,51.3,88.7,56.8z};
  }
}

\newcommand\orcidicon[1]{\href{https://orcid.org/#1}{\mbox{\scalerel*{
\begin{tikzpicture}[yscale=-1,transform shape]
\pic{orcidlogo};
\end{tikzpicture}
}{|}}}}

\usepackage{hyperref}
\usepackage{stmaryrd}
\usepackage[utf8]{inputenc}
\usepackage{rotating}
\usepackage{comment}
\usepackage{url}
\usepackage{bbm}
\usepackage{natbib}
\usepackage{soul}
\usepackage[normalem]{ulem}
\usepackage{tabularx,booktabs}
\usepackage{stmaryrd}
\newcolumntype{Y}{>{\centering\arraybackslash}X}

\begin{document}
\title{Closed-form solution of a general three-term recurrence relation: applications to Heun functions and social choice models}
\author{James Holehouse\orcidicon{0000-0002-3177-6069}}%
\email{jamesholehouse1@gmail.com}
\affiliation{The Santa Fe Institute}


\date{\today}

\begin{abstract}
    We derive a concise closed-form solution for a linear three-term recurrence relation. Such recurrence relations are very common in the quantitative sciences, and describe finite difference schemes, solutions to problems in Markov processes and quantum mechanics, and coefficients in the series expansion of Heun functions and other higher-order functions. Our solution avoids the usage of continued fractions and relies on a linear algebraic approach that makes use of the properties of lower-triangular and tridiagonal matrices, allowing one to express the terms in the recurrence relation in closed-form in terms of a finite set of orthogonal polynomials. We pay particular focus to the power series coefficients of Heun functions, which are often found as solutions in eigenfunction problems in quantum mechanics and general relativity and have also been found to describe time-dependent dynamics in both biology and economics. Finally, we apply our results to find equations describing the relaxation times to steady state behaviour in social choice models.
\end{abstract}

\maketitle

\footnotesize \noindent \textbf{Disclaimer: This article was published in \cite{holehouse2023recurrence} under a more applied narrative. However, I have kept this preprint separate, as I believe provides a useful contribution to three-term recurrence relations outside of applied narratives. The original title I chose, ``\textit{Closed-form solution of a general three-term recurrence relation $\ldots$}'', is a little misleading, since the usage of ``closed-form'' is not quite correct\footnote{Closed-form solutions are not permitted iterative aspects, i.e., as seen in infinite continued fractions.}. Nevertheless, this manuscript: (i) provides an up-to-date literature review on three-term recurrence research; (ii) makes it clear that continued fraction solutions of three-term recurrence relations (e.g., see \cite{risken1980solutions}) are equivalent to iterative solutions composed of orthogonal polynomials (see Eqs.~\eqref{eq:orthogPolys}-\eqref{eq:orthogPolys2}); (iii) clarifies that the general solution to a three-term recurrence relation is a Heun function, whose main difficulty lies in not having closed-form series coefficients in its series expansion; and (iv) emphasizes that the difficulty of eigenfunction solutions to master equations that admit Heun eigenfunctions is in the lack of a closed-form solution for the eigenvalues (see Eq.~\eqref{eq:finiteCF}). With this disclaimer in mind, I have not changed the main text of the article. For those further interested in this topic, I urge you to read Gautschi's classic \cite{gautschi1967computational} as well as Risken's work on solutions to three-term recurrence relations (but also higher-order recurrences) via continued fractions \cite{risken1980solutions}.}

\normalsize

\section{Introduction}


Recurrence relations are ubiquitous in quantitative science. They describe finite difference schemes necessary to compute derivatives in partial and ordinary differential equations, master equations and backward equations commonly used to model stochastic dynamics \cite{van1992stochastic,gardiner2009stochastic}, the coefficients in the power series of special functions \cite{dlmf2017digital}, and share some common concepts with continued fractions \cite{gautschi1967computational}. Often, they arise in applied studies of Markov processes in the biological literature, e.g., enzyme kinetics \cite{grima2017exact,schienbein1997enzyme}, gene expression \cite{szavits2022mean,ham2020exactly}, molecular motors \cite{antal2005burnt}, and asymmetric exclusion processes \cite{nossan2013disordered,blythe2007nonequilibrium}. Common methods to solve them include iteration, ansatzes, and generating function approaches \cite{accikgoz2010generating}. However, each of these methods only works in limited special cases. For example, two-term recurrence relations (solved by iteration) or recurrence relations of arbitrary order with constant coefficients (solved by an exponential ansatz giving a characteristic equation) have well-known solutions. 
The general three-term recurrence relation that we consider in this paper is,
\begin{align}\label{eq:genRR}
    R_j C_{j+1}-\tilde{Q}_j C_j+P_j C_{j-1} = 0,\; j\in\{0,1,2,\dots\},
\end{align}
with the boundary conditions $C_0=1$ and $C_{-1}=0$, and where the coefficients $R_j$, $\tilde{Q}_j$ and $P_j$ have a general dependence on $j$. 

Although commonly stated as being unsolved (e.g., see the section on Heun functions in the handbook of Maple \cite{maple}), some papers in the past decade have made progress in solving three-term recurrence relations. Recent work by Choun \cite{choun2013analytic}, from a series of studies that include \cite{choun2012generalization,choun2013special}, tackles the problem of solving the three-term recurrence relation defining the Heun function and looks to find the conditions under which Heun functions reduce to finite polynomials. Unfortunately, the proposed solution is difficult to verify and unwieldy (see \cite[Eq.~(5)]{choun2013analytic})\footnote{Additionally, most of this series of work, aside from \cite{choun2013analytic}, remains unpublished.}. Similar conclusions can be made regarding another solution to the general three-term recurrence relation by Gonoskov \cite{gonoskov2014closed}, wherein the author defines and utilises \textit{recursive sum theory} (see \cite[Eq.~(47)--(48)]{gonoskov2014closed}). However, other approaches with greater applicability are found in the seminal work of Risken \textit{et al.~}\cite{risken1980solutions,risken1996fokker}, who study generalised recurrence relations, often with applications to Fokker-Planck equations, and solve them using continued fractions. Work of Haag \textit{et al.~}takes a similar approach \cite{haag1979exact}, and it is shown how exact solutions to the one-dimensional master equation can be found in terms of continued fractions (a work that precedes the cited work of Risken).

In this paper, we solve a general three-term recurrence relation using a simple linear algebraic method reliant on analytic results from the inversion of tridiagonal matrices \cite{usmani1994inversion}. This leads to expressions for the sequence $C_j$ in terms of determinants of tridiagonal matrix, which can be conveniently expressed in terms of products of orthogonal polynomials. These expressions allow one to see the analytic structure of the $C_j$ in terms of well-known mathematical operations.

An application of recurrence relations of particular importance is in providing closed-form expressions for the Frobenius solutions of higher-order functions, whose coefficients in a series expansion are described by three-term (or higher-order) recurrence relations. These higher-order functions have been shown to be particularly relevant in the solutions of the master equations describing non-trivial models of binary choice \cite{holehouse2022exact} and community assembly \cite{mckane2000mean}. By \textit{higher-order} we mean that the number of singularities defining the function is greater than the number defining the hypergeometric differential equation (i.e., more than three), in which case the coefficients in the series expansion satisfy a two-term recurrence relation and can be solved by either Pochhammer or gamma functions \cite[Chap.~15]{dlmf2017digital}. The next highest order Fuchsian differential equation with four regular singularities defines the \textit{general Heun function}, whose Frobenius solutions satisfy a three-term recurrence relation. Due to the increasing complexity of problems considered in the physics literature, Heun functions are becoming increasingly common and describe solutions to problems in quantum mechanics \cite{el2008solutions,ralko2002heun,manning1935energy}, general relativity \cite{el2008solutions,leaver1986solutions} and have some applications to stochastic processes \cite{jain2020evolutionary} (see the review of Hortaccsu \cite{hortaccsu2018heun} and the references therein for further examples). A closed-form derivation of the series coefficients in the Frobenius solutions of Heun functions would allow researchers to easily obtain expressions defining polynomial solutions to Heun's differential equation, and in the process determine the relaxation spectra of non-trivial social choice models.

The structure of this paper is as follows. In Sec.~\ref{sec:CFsol} we solve the recurrence relation in Eq.~\eqref{eq:genRR} using linear algebraic methods leading to the main result of our paper given by Eq.~\eqref{eq:mainRes}, and we relate this solution to the previous work conducted by Risken \cite{risken1996fokker} in Section \ref{sec:contF}. Then, in Sec.~\ref{sec:HeunFns} we review both Heun's general and confluent differential equations, and show how our general solution solves for the coefficients in the Frobenius solutions in closed-form. In Sec.~\ref{sec:socialChoice} we show how our results allow one to determine the relaxation rates to equilibrium in two models of social choice that have eigenfunctions described by functions of whose Frobenius solutions satisfy the three-term recurrence relation. Finally in Sec.~\ref{sec:conc} we conclude the study.

\section{Closed-form solution of a three-term recurrence relation}\label{sec:CFsol}
We begin by re-writing the three-term recurrence relation in Eq.~\eqref{eq:genRR} as a matrix equation. First we define,
\begin{align}
    \mathbb{A}=\begin{pmatrix}
        R_0 & & & & \\
        -\Tilde{Q}_1 & R_1 & & & \\
        P_2 & -\Tilde{Q}_2 & R_2 & \\
        0 & P_3 & -\Tilde{Q}_3 & R_3 & \\
         \vdots & & & & \ddots
    \end{pmatrix}
\end{align}
where  $\mathbb{A}$ is a square infinite-dimensional lower-triangular matrix. Note that because $\mathbb{A}$ is lower-triangular the eigenvalues of $\mathbb{A}$ are $R_i$ for $i\in\{0,1,2,\ldots\}$. Then the recurrence relation in Eq.~\eqref{eq:genRR} is equivalent to the following,
\begin{align}\label{eq:triang}
    \mathbb{A}\cdot \vec{C} = \vec{l},
\end{align}
where we have defined the infinite-dimensional column vectors,
\begin{align}
    \vec{C} = \begin{pmatrix}
        C_1\\
        C_2\\
        C_3\\
        C_4\\
        \vdots
    \end{pmatrix},\;
    \vec{l} = \begin{pmatrix}
        \tilde{Q}_0\\
        -P_1\\
        0\\
        0\\
        \vdots
    \end{pmatrix},
\end{align}
where the only two non-zero elements of $\vec{l}$ are $l_1$ and $l_2$.
Then $\vec{C}$ is given by,
\begin{align}\label{eq:coeffs1}
    \vec{C} = \mathbb{A}^{-1}\cdot \vec{l}.
\end{align}
Therefore, if we can find the inverse of $\mathbb{A}$ then we have solved for the general three-term recurrence relation in Eq.~\eqref{eq:genRR}. In the following, we denote the inverse elements of $\mathbb{A}$ as $\theta_{i,j}\equiv [\mathbb{A}^{-1}]_{i,j}$, and carrying out the multiplication in Eq.~\eqref{eq:coeffs1} we find,
\begin{align}
    C_i = \tilde{Q}_0\theta_{i,1}-P_1\theta_{i,2}.
\end{align}
Hence there are two sets of inverse elements that we require, those in the first and second columns of $\mathbb{A}^{-1}$. To find the matrix inverse we make use of Cramer's rule \cite{higham2002accuracy},
\begin{align}\label{eq:Cramers}
    \theta_{i,j} = \frac{(-1)^{i+j}M_{j,i}}{\textrm{det}(\mathbb{A})},
\end{align}
where $M_{j,i}$ is a minor of $\mathbb{A}$, i.e., the determinant of $\mathbb{A}$ with row $j$ and column $i$ removed, and the determinant of $\mathbb{A}$ is simply the product of the eigenvalues of $\mathbb{A}$ that is given by,
\begin{align}\label{eq:det}
    \textrm{det}(\mathbb{A}) = \prod_{i=0}^\infty R_i.
\end{align}
Although formally the determinant of an infinite matrix is not well-defined, we show in the Appendix that one does not have to evaluate this infinite product as cancellation occurs with the minors $M_{j,i}$ in the numerator of $\theta_{i,j}$. The remaining task is then to find the minors $M_{1,i}$ and $M_{2,i}$. In Appendices \ref{sec:Mi1} and \ref{sec:Mi2} we find expressions for $M_{1,i}$ and $M_{2,i}$ explicitly in terms of a sequence of polynomials in $x$,
\begin{align}\label{eq:orthogPolys}
    \phi_0^i(x) &= 1,\; \phi_1^i(x) = x+\Tilde{Q}_i,\;\\\label{eq:orthogPolys2}
    \phi_j^i(x) &= (\Tilde{Q}_{i-(j-1)}+x)\phi^i_{j-1}(x)-R_{i-(j-1)}P_{i-(j-2)}\phi_{j-2}^i(x),\;j\in\{2,3,\ldots,i-1,i\}.
\end{align}
where the superscript $i$ (plus 1) gives the number of polynomials in the set, and the subscript $j$ denotes the $(j+1)$-th recursively defined polynomial. This result was first shown by \cite{usmani1994inversion}, and allows one to express the determinant of a tridiagonal matrix in an analytic and computationally convenient way. Note, that the polynomials $\phi_j^i(x)$ are orthogonal following Shohat--Favard theorem \cite{chihara2011introduction}, although we do not use this property directly. Then we can express $\theta_{i,1}$ and $\theta_{i,2}$ as,
\begin{align}
    \theta_{i,1} &= \frac{\phi^{i-1}_{i-1}(0)}{\prod_{j=0}^{i-1}R_j},\\
    \theta_{i,2} &= \frac{\phi^{i-1}_{i-2}(0)}{\prod_{j=1}^{i-1}R_j}.
\end{align}
This gives us our closed-form expression for $C_i$ in terms of the recursively defined orthogonal polynomials $\phi_j^i(x)$,
\begin{align}\label{eq:mainRes}
    C_i = \frac{\tilde{Q}_0 \phi^{i-1}_{i-1}(0) - R_0 P_1 \phi^{i-1}_{i-2}(0)}{\prod_{j=0}^{i-1}R_j},
\end{align}
which is the main result of the paper. It elucidates the functional dependence of $C_i$ on the coefficients $P_j,Q_j$ and $R_j$ in the recurrence relation itself. Note that this result is much more compact, and its derivation much easier, than other solutions to three-term recurrence relations described in \cite{choun2013analytic,gonoskov2014closed}. 
It also avoid restrictions on the $C_i$, such as those imposed in the solution of Risken \cite{risken1996fokker}, wherein the authors require that for some $N$ large enough that $C_{N+1}=0$. 
We will see in the final section of the paper that this result allows us to easily find the conditions under which polynomial solutions to Heun's differential equations, or indeed any function defined by a three-term recurrence relation, occur. But first, we establish the connection between Eq.~\eqref{eq:mainRes} and the work of Risken \textit{et al.} \cite{risken1980solutions,risken1996fokker}.

\section{Relationship to continued fractions}\label{sec:contF}
As stated in the introduction, the most useful previous solutions to three-term recurrence come in the form of continued fractions which are more cumbersome than the results derived in the previous Section. Here we show the relationship between our solution and that of \cite{risken1980solutions}, showing how the coefficients $C_i$ are equivalently given by a finite product over a set of continued fractions.

In \cite{risken1980solutions}, Risken and Vollmer study non-stationary recurrence relations of the scalar and vector type. The scalar type is defined by,
\begin{align}\label{eq:Risken1}
    \frac{d C_j(t)}{d t} = R_j C_{j+1}(t)-\tilde{Q}_j C_j(t)+P_j C_{j-1}(t),\; j\in\{0,1,2,\dots\},
\end{align}
where for some finite value of $N$, $C_{N+1}(t)=0$, which can either be an artificial truncation leading to approximate $C_j(t)$, or exact in some special cases. Such time-dependent recurrence relations are common in the study of one-dimensional master equations and first-passage time processes \cite{van1992stochastic,gardiner2009stochastic,smith2015general,ashcroft2015mean,barrio2013reduction}. To solve Eq.~\eqref{eq:Risken1}, one can treat it as an initial value problem (as was also done in \cite{haag1979exact}), but the easiest way to solve it is as an eigenvalue problem. To do this one makes the separation ansatz $C_j(t) = \tilde{C}_je^{-\lambda t}$, which leads to a homogeneous three-term recurrence relation of the type seen in Eq.~\eqref{eq:genRR}, explicitly,
\begin{align}\label{eq:Risken2}
    R_j \tilde{C}_{j+1}-(\tilde{Q}_j-\lambda) \tilde{C}_j+P_j \tilde{C}_{j-1} = 0.
\end{align}
The problem of finding $C_j(t)$ is then split into two. First, one needs to find the expression governing the $\tilde{C}_{j}$. Second, one must find the eigenvalues $\lambda$. Finding the $\lambda$'s is a classic problem of linear algebra, and amounts to finding the values of $\lambda$ under which the following holds,
\begin{align}
    \begin{vmatrix}
        (\lambda - \Tilde{Q}_0) & R_0 & & & & \\
        P_1 & (\lambda - \Tilde{Q}_1) & R_1 & & &  \\
        & P_2& (\lambda - \Tilde{Q}_2) & & & & \\
        & & & \ddots & & &\\
        & & & & (\lambda - \Tilde{Q}_{N-1}) & R_{N-1} & \\
        & & & & P_N&  (\lambda - \Tilde{Q}_N) &
    \end{vmatrix} = 0,
\end{align}
for which there will be $N+1$ solutions for $\lambda$ where $\tilde{C}_{N+1}=0$. Generally, this can be done numerically. Clearly, usage of the exponential ansatz reduces the problem of calculating the $\tilde{C}_{j}$ to the same problem as we initially consider in Eq.~\eqref{eq:genRR}, but Risken \textit{et al.~}\cite{risken1980solutions,risken1996fokker} solve it using continued fraction methods, as opposed to the method of orthogonal polynomials introduced above. Consider the transformation $S_j = \tilde{C}_{j+1}/\tilde{C}_j$, which transforms Eq.~\eqref{eq:Risken2} into,
\begin{align}
    R_j S_j +(\lambda-\Tilde{Q}_j)+\frac{P_j}{S_{j-1}} = 0,
\end{align}
which can be solved for $S_j$ to give the recursive relationship,
\begin{align}
    S_j = \frac{-P_{j+1}}{(\lambda-\tilde{Q}_{j+1})+R_{j+1}S_{j+1}}.
\end{align}
This relationship can be iterated to give an expression for $S_j$ in terms of a continued fraction,
\begin{align}
    S_j = \frac{-P_{j+1}}{\lambda-\tilde{Q}_{j+1}-}\frac{R_{j+1}P_{j+2}}{\lambda-\tilde{Q}_{j+2}-}\ldots \frac{R_{N-1}P_{N}}{\lambda-\tilde{Q}_{N}},\; j\in\{1,2,\dots,N-2\},
\end{align}
with $S_N = 0$ and $S_{N-1}= -P_N/(\lambda-\Tilde{Q}_N)$. One can then recover the $\tilde{C}_j$ by noticing that,
\begin{align}\label{eq:BC}
    \tilde{C}_j = \tilde{C}_0\prod_{i=0}^{j-1}S_i.
\end{align}
From here it is clear that Eq.~\eqref{eq:mainRes} is equivalent to a finite product over a set of continued fractions. The benefit of the result in Eq.~\eqref{eq:mainRes} is that it is valid even for non-physical recurrence relations that grow unboundedly, and there is no restriction that $C_{N+1}(t) = 0$ for some values of $N$. The results of Risken \textit{et al.~}for scalar three-term recurrence can hence be seen as a special case of Eq.~\eqref{eq:mainRes}. Of course, for many physical applications the recurrence does not grow unboundedly, as often the recurrence variable represents physical variables or probabilities. This however is not a restriction on the special functions considered in the next section.

\section{Heun functions}\label{sec:HeunFns}

\begin{figure}[h!]
    \includegraphics[width=0.8\textwidth]{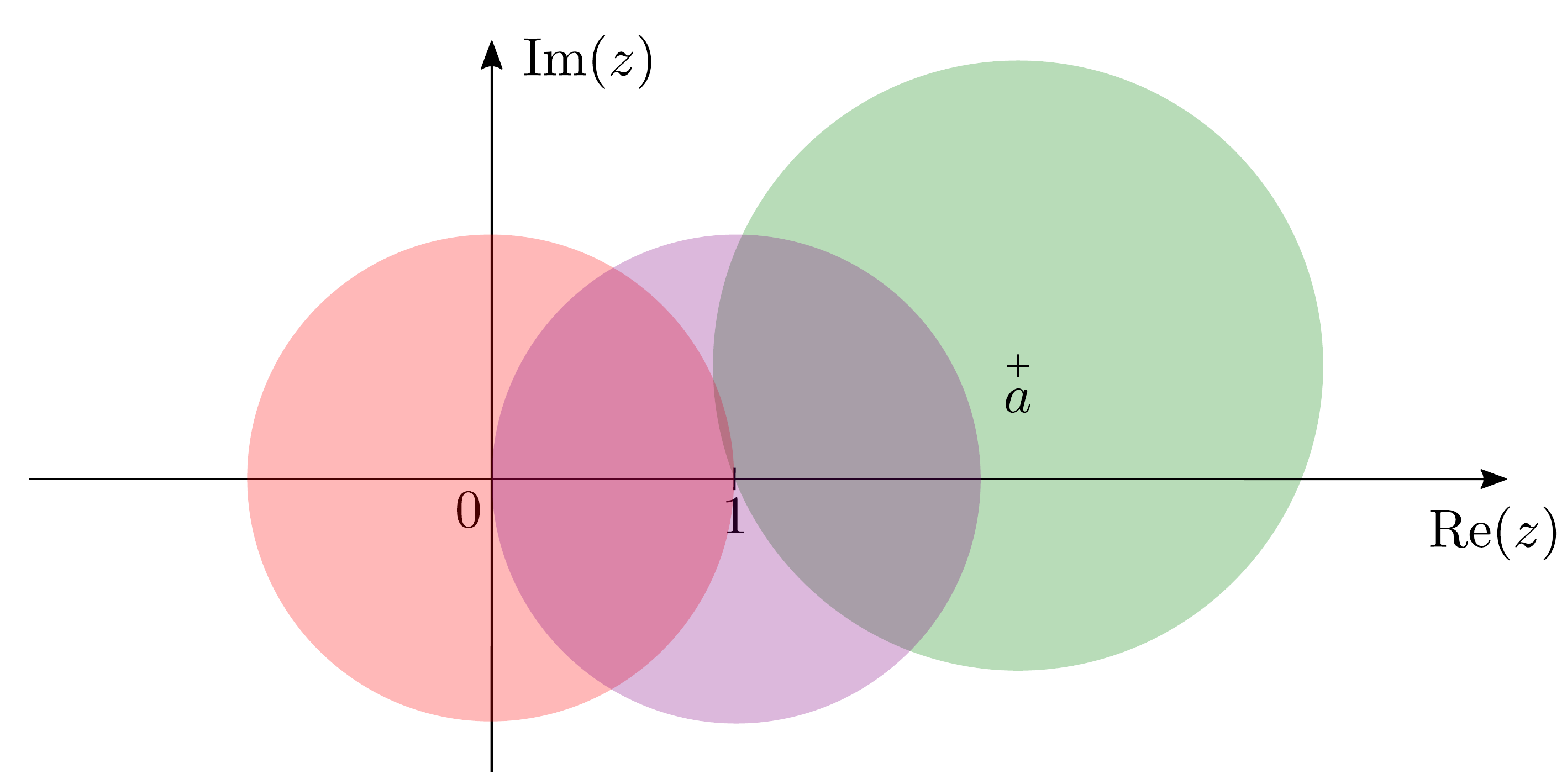}
    \caption{Illustration showing the radii of convergence of the Frobenius solutions of the general Heun equation about $z=0,1$ and $a$. The parameter $a$ can take any complex value. Each Frobenius solution is valid in a circle in the complex plane (centred at the respective singularity) valid until the next singularity. Ordinary series expansions, not about singularities, are also valid in a circle extending up to the next singularity. Note that in cases where the Heun functions simplify to polynomials, or even Frobenius solutions valid at two singularities, the radii of convergence will extend beyond those in the illustration. The radius of convergence of the solution at $z=\infty$ can be seen clearly through the independent variable transformation $z\to1/x$, as shown in \cite[p.~15]{ronveaux1995heun}. For further details see \cite{motygin2015numerical}.}
    \label{fig1}
\end{figure}

Heun functions have had increased popularity in the study of Markov processes as researchers attempt to make their models more general, and eigenfunctions of the master equation or Fokker-Planck equations they consider can no longer be described by hypergeometric or lesser-order functions. They are the solution to a differential equation with four regular singularities, given by the ODE \cite{ronveaux1995heun,ince1956ordinary},
\begin{align}\label{eq:ha1}
    \frac{d^2 y}{d z^2}+\Big(\frac{\gamma}{z}+\frac{\delta}{z-1}+\frac{\epsilon}{z-a}\Big)\frac{d y}{d z}+\frac{\alpha \beta z - q}{z(z-1)(z-a)}y=0,
\end{align}
whose singularities are at $z=0,1,a$ and $\infty$, around each of which one can form a Frobenius solution of two linearly independent general Heun functions. To ensure that the Frobenius indices at $z=\infty$ are $\{\alpha, \beta\}$, the relation $\alpha+\beta+1=\gamma+\delta+\epsilon$ must be satisfied. This ODE is known as \textit{Heun's general equation}, and it is the natural extension of the hypergeometric differential equation \cite[Sec.~15.2]{dlmf2017digital}, being a second-order linear Fuchsian equation with four singularities. For clarity, we show the associated radii of convergence of each Frobenius solution of the general Heun equation in Fig.~\ref{fig1}. Confluent forms of the Heun function also arise through limits of the solution to Heun's general equation. For example, merging the singularities of the general Heun function at $z=a$ and $z=\infty$, by taking $a\to \infty$ and simultaneously $q,\alpha\beta,\epsilon\to \infty$ in such a way that $\epsilon/a \to -\epsilon'$, $q/a\to q'$ and $\alpha\beta/a\to \alpha'$, one arrives at the confluent Heun equation,
\begin{align}\label{eq:ha6}
    \frac{d^2 y}{d z^2}+\Big(\epsilon'+\frac{\gamma}{z}+\frac{\delta}{z-1} \Big)\frac{d y}{d z} + \frac{\alpha' z - q'}{z(z-1)}y=0,
\end{align}
which has two regular singularities at $z=0,1$ and an irregular singularity of rank 1 at $z=\infty$. 
In what follows we simply relabel the parameters of the confluent Heun function by dropping the prime, i.e., $\epsilon'\to\epsilon,\alpha'\to\alpha$ and $q'\to q$. One can then merge the singularities in the confluent Heun equation to derive further confluent types of Heun functions \cite[Sec.~31.12]{dlmf2017digital}. The Frobenius indices for each of Eqs.~\eqref{eq:ha1} and \eqref{eq:ha6} around each regular singularity are well reported \cite{ronveaux1995heun,dlmf2017digital}, and here we consider series solutions to Eqs.~\eqref{eq:ha1} and \eqref{eq:ha6} with the Frobenius index of $0$\footnote{Note that our results below can be trivially applied to the solution defined by the second Frobenius exponent at each singularity, but with re-defined parameters $P_j$, $Q_j$ and $R_j$.}. This means assuming that $y(z)$ around $z=0$ has the form,
\begin{align}\label{eq:ps}
    y(z) = \sum_{j=0}^\infty C_j z^j.
\end{align}
Substituting this into Eqs.~\eqref{eq:ha1} and \eqref{eq:ha6} results in the three-term recurrence relation defining the solution at $z=0$ with the Frobenius index of 0,
\begin{align}\label{eq:HeunRR}
    R_j C_{j+1}-(Q_j+q)C_j+P_j C_{j-1} = 0,\; j\in\{0,1,2,\dots\},
\end{align}
with the boundary conditions $C_0=1$ and $C_{-1}=0$. The coefficients in the recurrence relation are different for the general and confluent Heun equations, and can be derived through the standard substitution of Eq.~\eqref{eq:ps} into the respective Heun differential equation. For the general Heun equation in Eq.~\eqref{eq:ha1} we have,
\begin{align}\nonumber
    P_j &= (j-1+\alpha)(j-1+\beta),\\\nonumber
    Q_j &= j\left( (j-1+\gamma)(1+a)+a \delta+\epsilon \right),\\\nonumber
    R_j &= a(j+1)(j+\gamma),
\end{align}
whereas for the confluent Heun equation we have,
\begin{align}\nonumber
    P_j &=  (1-j)\epsilon-\alpha,\\\nonumber
    Q_j &=  j(1-j)+j(\epsilon-\gamma-\delta),\\\nonumber
    R_j &=  (j+1)(j+\gamma).
\end{align}
Clearly, for functions of the Heun class, Eq.~\eqref{eq:HeunRR} is essentially Eq.~\eqref{eq:genRR} but with $\Tilde{Q}_j = Q_j +q$, meaning that the $C_j$ can be solved directly by a slight modification to Eq.~\eqref{eq:mainRes},
\begin{align}\label{eq:heunCoeffs}
    C_i = \frac{q \phi^{i-1}_{i-1}(q) - R_0 P_1 \phi^{i-1}_{i-2}(q)}{\prod_{k=0}^{i-1}R_k},
\end{align}
where the orthogonal polynomials are now evaluated at the accessory parameter $x=q$. This result shows explicitly why we must have $\gamma\notin\{0,-1,-2,\ldots\}$, since this would lead to a zero in the denominator of $C_i$. Note that in cases where the parameters of the Heun functions are such that they reduce to polynomials, or to solutions valid at more than one singularity, the $C_i$ will consist of a convergent series as $i\to\infty$, even outside their standard radius of convergence \cite{ronveaux1995heun}. However, in general the results of Risken do not apply for general Heun or confluent Heun functions outside the standard radius of convergence \cite{risken1980solutions}.


\section{Application to relaxation times in models of social choice}\label{sec:socialChoice}
In this section we apply the above analytics to explore the relaxation times to equilibrium in two distinct models of social choice, modelled as continuous-time Markov processes. Using the generating function approach to the master equation, one can show that the eigenspectra that defines the time-dependence in the dynamics can be found through imposing physical restrictions on the generating function, which accounts for the finite size of the agent populations. This also allows one to connect continued fractions to their equivalent polynomial expressions that define the eigenspectra. The key references for the examples in this section are \cite{kirman1993ants,lambiotte2007dynamics,holehouse2022exact}.

\subsection{Fully asymmetric binary choice model}
Many social choices are well-described by binary choice situations wherein a fixed number of $N$ agents decide between a \textit{left} choice and a \textit{right} choice with respect to two influences, (1) a random switching of decision of each agent, and (2) a recruitment whereby agents deciding one way can recruit other to the same decision. Such models have been increasingly common as they are much more analytically tractable than multiple choice scenarios \cite{borghesi2007songs}, and many social decisions can be approximated as being for-or-against a specific choice (even in a multiple choice scenario). This situation describes the model of ant recruitment used by Kirman \cite{kirman1993ants} to show how endogenous interactions can induce polarity in the collective decisions of agents and that polarity does not necessarily require an exogenous force. The same model has been used in other contexts to describe genetic drift \cite{moran1958random} and the dynamics of migration \cite{mckane2000mean,mckane2004analytic}. In simpler cases where the effects of recruitment are symmetric in both decisions the binary choice model has been solved \cite{mckane2000mean,holehouse2022exact}. However, making the effects of recruitment asymmetric leads to non-trivial relaxation rates and eigenfunctions for the stochastic process \cite{holehouse2022exact}. The fully asymmetric system defining the binary choice model is given by,
\begin{equation}\label{eq:reac_fullasym}
L \xrightleftharpoons[n\varepsilon_2+n(N-n)\mu_2]{(N-n)\varepsilon_1 +n(N-n)\mu_1} R,
\end{equation}
where the expressions above and below the arrows indicate the propensity (per agent per unit of time) for a reaction to occur (determined from mass-action kinetics \cite{schnoerr2017approximation}), $L$ and $R$ denote an agent deciding left or right respectively, and it is assumed that each agent is equally likely to interact with any other, meaning network effects can be ignored \cite{mckane2000mean}. $\varepsilon_1$ and $\varepsilon_2$ are the random switching rates from left-to-right and right-to-left respectively, and $\mu_1$ and $\mu_2$ are the respective rates of recruitment. In the propensities $n$ denotes the number of agents deciding \textit{right} meaning that $(N-n)$ agents decide \textit{left}. Note that this is a second-order reaction scheme due to the interactions between $L$ and $R$ agents. 

This reaction scheme corresponds to the master equation,
\begin{equation}\label{eq:ASCme}
\begin{split}
\partial_t P(n,t) = &\left[(N-(n-1))\varepsilon_1 +  \mu_1(n-1)(N-(n-1))\right]P(n-1,t)\\&+ \left[(n+1)\varepsilon_2 +  \mu_2(n+1)(N-(n+1))\right]P(n+1,t)\\
&-\left[(N-n)\varepsilon_1 +n\varepsilon_2+(\mu_1+\mu_2)n(N-n)\right]P(n,t),
\end{split}
\end{equation}
where $P(n,t)$ is the probability of observing $n$ right-deciding agents at a time $t$. The standard next step is to introduce the generating function $G(z,t)=\sum_n P(n,t)z^n$ which converts the master equation (a set of coupled first-order ODEs) into a single PDE, which we give in Appendix \ref{sec:appC1}. Using separation of variables one can show that $G(z,t)\sim f_\lambda(z) e^{-\lambda t}$, and the PDE defining $G(z,t)$ reduces to a second-order ODE in $f_\lambda(z)$ whose solution is a general Heun function (also see \cite{holehouse2022exact}),
\begin{align}\label{eq:genGenFn}
    f_\lambda(z) = H(a,q(\lambda);\alpha,\beta,\gamma,0;z),
\end{align}
where we have defined,
\begin{equation}\label{eq:heundefs}
\begin{split}
    a &= \mu_2/\mu_1,\\
    q(\lambda) &= \frac{(\lambda-N\varepsilon_1)(N-1)}{\mu_1},\\
    \alpha &= -N,\\
    \beta &= \frac{(N-1)\varepsilon_1}{\mu_1},\\
    \gamma &= -(N-1)\left(1+\frac{\varepsilon_2}{\mu_2}\right),
    \end{split}
\end{equation}
where the parameters have the same meaning as introduced for the general Heun function in Section \ref{sec:HeunFns}. Note that $\delta=0$. In order for the $f_\lambda(z)$ to be physical we require that the $\lambda$ be chosen such that $f_\lambda(z)$ is a polynomial of order $N$ in $z$. This amounts to choosing $\lambda$ such that $P(N+1,t)=0$, for which we can easily find a polynomial defining this from Eq.~\eqref{eq:heunCoeffs},
\begin{align}\label{eq:ESeq}
    P(N+1,t) \propto q(\lambda) \phi^{N}_{N}(q(\lambda)) - R_0 P_1 \phi^{N}_{N-1}(q(\lambda))=0,
\end{align}
which is a polynomial in $\lambda$ of order $N+1$, whose roots define the eigenspectrum of relaxation to the equilibrium state. One can then additionally show the equivalence between the finite continued fractions and the polynomials defining the eigenspectrum, where using the formula in terms of continued fractions in \cite[Eq.~(28)]{holehouse2022exact} and equating the $q(\lambda)$ terms in each expression one finds,
\begin{align}\label{eq:finiteCF}
    \frac{1}{Q_1+q-}\frac{R_1P_2}{Q_2+q-}\dots \frac{R_{N-1}P_N}{Q_N+q} = \frac{\phi^{N}_{N-1}(q)}{\phi^{N}_{N}(q)}.
\end{align}
This allows one to easily find the rational fraction corresponding to a continued fraction of the above form in terms of orthogonal polynomials in $q$. This holds for any $P_j,Q_j$ or $R_j$ for which there is some $C_{N+1}=0$.


\subsection{The vacillating voter model}
We can also use our method to easily derive polynomials describing the eigenspectra of models whose eigenfunctions satisfy generating function ODEs that are more complex than Heun functions, as long as the special functions defining them have series expansions whose coefficients are described by a three-term recurrence. For example, consider the following third-order reaction scheme that describes so-called vacillating voters \cite{lambiotte2007dynamics},
\begin{equation}\label{eq:voter_reaction}
    L\xrightleftharpoons[p_d n +(1-p_d) \frac{(N-n)n}{N-1}\left(1+\frac{n}{N-1}\right)]{p_d (N-n) +(1-p_d) \frac{(N-n)n}{N-1}\left(1+\frac{N-n}{N-1}\right)}R,
\end{equation}
where $L$ and $R$ again correspond to two different decisions, but now with different dynamical rules as compared to the asymmetric binary choice model. The model was solved semi-analytically in \cite{holehouse2022exact}. The rules described by this process are as follows. An agent is chosen at random from the population, and with probability $p_d$ changes their decision. However, with probability $(1-p_d)$ the agent looks at the decision of another agent. If this agent agrees with the originally chosen agent nothing happens, but if there is a disagreement the original agent will then select another agent at random and perform the same procedure again. Only if both other agents selected by the original agent disagree with their current view will the original agent change their mind. As one might expect, this leads to quite different behaviours from the original voter model \cite{liggett1999stochastic}, including transient and steady state trimodality \cite{lambiotte2007dynamics,holehouse2022exact}.

Again, one can construct a master equation describing the dynamics of the vacillating voters and can write the corresponding generating function equation. In Appendix \ref{sec:appC2} we show this, and again use separation of variables to define the equation which $f_\lambda(z)$ satisfies, which is a third-order ODE in $f_\lambda(z)$ and the unspecified spectral parameter $\lambda$. Employing a series solution about $z=0$, i.e., $f_\lambda(z) = \sum_j C_jz^j$, one then finds the following recursion relation for the coefficients $C_{j}$,
\begin{equation}\label{eq:rec}
\begin{split}
&C_0=1,\quad(N-1)((N-1)p_d+N(1-p_d)) C_1 -q(\lambda)C_0=0,\\
&R_j C_{j+1}-(Q_j+q(\lambda))C_j+P_j C_{j-1} = 0,
\end{split}
\end{equation}
with the condition that $C_{N+1}=0$, and where we re-define,
\begin{equation}\label{eq:rec_Defs}
\begin{split}
q(\lambda) =& (N-1)(p_d N-\lambda),\\
R_j =& (j+1) \bigg(j \bigg(-j^2+j-2\bigg) (1-p_d)\\
& +(1-p_d)  N (N-1)+(N-1)^2 p_d \bigg),\\
Q_j =& -j (1-p_d)  (3 N-2) (N-j),\\
P_j =& (j-1) (1-p_d)  (j-2 N) (j-N-1)\\
&-(N-1) p_d  ((j-2) N-j+1),
\end{split}
\end{equation}
which has been taken directly from \cite{holehouse2022exact}. Again the polynomial describing the eigenspectra will be given by Eq.~\eqref{eq:ESeq} but now with the redefined $q(\lambda),R_j,Q_j$ and $P_j$.

\section{Discussion}\label{sec:conc}


In this paper we have provided a closed-form solution to a general three-term recurrence relation that determines the relaxation spectra in non-trivial models of binary choice. This allowed us to express the sequence defined by the recurrence in terms of orthogonal polynomials that allow one to easily see the analytic structure of terms in the sequence. Our solution is not reliant on the convergence of the recurrence, unlike that of the continued fraction solution, meaning that it can be applied even in situations where the sequence defined by the recurrence grows unboundedly. We then showed how this result provides the series coefficients in the Frobenius expansions of Heun functions. In the final Section we used these analytics for Heun functions, and other special functions whose Frobenius solutions satisfy a three-term recurrence, to derive concise polynomial expressions that define the eigenspectra for relaxation to the steady state in two distinct models of social choice.


Our result has clear analytic use, e.g., easily computing the polynomial satisfied by the eigenspectrum of a continuous-time Markov process (Section \ref{sec:socialChoice}) or expressing finite continued fractions as a rational fraction (Eq.~\eqref{eq:finiteCF}). However, a computational limitation of our solution is that each $C_j$ will take the same order of time to compute as direct forward substitution on the triangular matrix equation in Eq.~\eqref{eq:triang}, although solving via this method does not lead to a closed-form solution (i.e., each $C_j$ would depend on all $C_{i<j}$ preceding it). We note that the same restriction applies to the continued fraction solution to the three-term recurrence relation provided by Risken \cite{risken1980solutions}. However,\textit{ it is often the analytical structure that is of interest to us in solving physical problems}---as we have shown in Section \ref{sec:socialChoice}.

Several avenues for further study remain open. The first is in the extension of the results presented herein to higher-order recurrence relations. Such an approach has been previously considered by Risken \cite{risken1980solutions}, wherein higher-order recurrence relations are converted into three-term \textit{vector recurrence relations} that can be solved by continued fraction methods very similar to those used for three-term scalar recurrence relations. Using a similar approach, it may be possible to generalise the results in this paper to higher-order recurrence relations in a way that does not require the usage of matrix continued fractions. The results that we have presented also allow for connections to be drawn to other parts of the Markov process literature involved in solving one-dimensional master equations for various problems, such as its time-dependent solution with arbitrary rates, or the one-dimensional first-passage time probabilities with absorbing \cite{ashcroft2015mean} and reflecting \cite{noskowicz1990first} boundaries \cite{redner2001guide}. These papers highlight the utility of studying three-term recurrence under different boundary and initial conditions, and the results that we have found possibly allow for a unification of the results found therein. Finally, and most optimistically, it may be possible to use our methods to derive time-dependent solutions to chemical reaction networks involving reactions of bimolecular form, and multi-step reactions, and provide an extension to the generalised solutions of monomolecular reaction systems provided by Jahnke \textit{et al.}~in \cite{jahnke2007solving} and the solution to the one-dimensional, linear, one-step master equation \cite{smith2015general}. Calculation of these results would rely on finding an appropriate representation of reaction schemes involving bimolecular reactions, possibly in the form of a vector recurrence relation, that then allows for linear algebraic methods to become computationally useful.

\section*{Acknowledgements}
The author would like to thank Ramon Grima for advice on the connection of the work herein to that of Risken\textit{ et al.~}\cite{risken1980solutions,risken1996fokker} and Haag \textit{et al.~}\cite{haag1979exact}, and Sidney Redner for stimulating discussions, including future applications of the work conducted herein. Additionally the author thanks Ramon Grima, Kaan \"{O}cal and Augustinas \v{S}ukys for providing key feedback on the earlier stages of this manuscript. This publication is based upon work that is supported by the National Science Foundation under Grant No.~DMR-1910736.

\bibliographystyle{unsrt}
\bibliography{biblio}

\pagebreak
\clearpage
\widetext
\begin{center}
\textbf{\large Appendix}
\end{center}
\appendix

\section{Calculation of minors $M_{i,1}$}\label{sec:Mi1}
We start by calculating the minors $M_{1,1}$, $M_{1,2}$ and $M_{1,3}$ before presenting the pattern for general $M_{1,i}$. For $M_{1,1}$ we trivially find that,
\begin{align}
    M_{1,1} = 
    \begin{vmatrix}
        R_1 & & & \\
        -\Tilde{Q}_2 & R_2 & \\
        P_3 & -\Tilde{Q}_3 & R_3 & \\
         \vdots & & & \ddots
    \end{vmatrix} = \prod_{i=1}^\infty R_i,
\end{align}
where even though the determinant of an infinite matrix is not formally well-defined, the result holds for our calculations below and in the main text. For $M_{1,2}$, we find that,
\begin{align}
    M_{1,2} = 
    \begin{vmatrix}
        -\Tilde{Q}_1 & & & \\
        P_2 & R_2 & \\
         & -\Tilde{Q}_3 & R_3 & \\
         \vdots & & & \ddots
    \end{vmatrix} = -\Tilde{Q}_1\prod_{i=2}^\infty R_i,
\end{align}
And for $M_{1,3}$,
\begin{align}
    M_{1,3} = 
    \begin{vmatrix}
        -\Tilde{Q}_1 & R_1 & & & \\
        P_2 & -\Tilde{Q}_2 & & \\
        & P_3 & R_3 & & \\
        &  & -\Tilde{Q}_4 & R_4 & \\
         \vdots & & & & \ddots
    \end{vmatrix} = \begin{vmatrix}
        \Tilde{Q}_1 & R_1\\
        P_2 & \Tilde{Q}_2 
    \end{vmatrix}\cdot
    \prod_{i=3}^\infty R_i,
\end{align}
where we have made use of Schur's formula \cite{zhang2006schur} for the determinants of block matrices, which states that if one has the block matrix,
\begin{align}
    \mathbf{E} = \begin{pmatrix}
        \mathbf{A} & \mathbf{B} \\
        \mathbf{C} & \mathbf{D}
    \end{pmatrix},
\end{align}
for invertible $\mathbf{A}$ and $\mathbf{D}$ we have, 
\begin{align}
    \mathrm{det}(\mathbf{E}) = \mathrm{det}(\mathbf{A})\cdot\mathrm{det}(\mathbf{D} - \mathbf{C}\cdot \mathbf{A}^{-1}\cdot \mathbf{B}),
\end{align}
which, when either $\mathbf{B}$ or $\mathbf{C}$ consist entirely of zeros, reduces to,
\begin{align}
    \mathrm{det}(\mathbf{E}) = \mathrm{det}(\mathbf{A})\cdot\mathrm{det}(\mathbf{D}).
\end{align}
We hence see the emergence of a pattern whereby the minors are the product of the determinant of a tridiagonal matrix multiplied by an product over $R_i$. By defining $D_i$ as the following determinant,
\begin{align}
    D_i\equiv \begin{vmatrix}
        \Tilde{Q}_1 & R_1 & & & \\
        P_2 & \Tilde{Q}_2 & & & \\
        & & \ddots & & \\
        & & & \Tilde{Q}_{i-1} & R_{i-1} & \\
        & & & P_i & \Tilde{Q}_{i}
    \end{vmatrix},
\end{align}
with $D_0=1$, we can then express the minors $M_{1,i}$,
\begin{align}
    M_{1,i} = D_{i-1}(-1)^{i-1}\prod_{j=i}^\infty R_j.
\end{align}
This formula recovers the cases already shown above for $i=1,2,3$ and can be shown to agree for any value of $i\in \mathbb{N}_1$. Usage of Cramer's rule from Eq.~\eqref{eq:Cramers} then gives us the elements $\theta_{i,1}$,
\begin{align}
    \theta_{i,1} = \frac{D_{i-1}}{\prod_{j=0}^{i-1}R_j}.
\end{align}
One can express the tridiagonal determinant defining $D_{i-1}$ in terms of recursively defined polynomials following the work of \cite{usmani1994inversion}. The set of polynomials $\phi_j^i(x)$ is defined by,
\begin{align}
    \phi_0^i(x) &= 1,\; \phi_1^i(x) = x+\Tilde{Q}_i,\;\\
    \phi_j^i(x) &= (\Tilde{Q}_{i-(j-1)}+x)\phi^i_{j-1}(x)-R_{i-(j-1)}P_{i-(j-2)}\phi_{j-2}^i(x),\;j\in\{2,3,\ldots,i-1,i\}.
\end{align}
This allows us to identify $D_i = \phi_i^i(0)$, and therefore we find the result contained in the main text,
\begin{align}
    \theta_{i,1} = \frac{\phi^{i-1}_{i-1}(0)}{\prod_{j=0}^{i-1}R_j}.
\end{align}

\section{Calculation of minors $M_{i,2}$}\label{sec:Mi2}
Similar to Appendix \ref{sec:Mi1}, we begin by calculating $M_{2,1}$, $M_{2,2}$, $M_{2,3}$ and $M_{2,4}$ and then identify the pattern for general $M_{2,i}$. First for $M_{2,1}$ we find,
\begin{align}
    M_{2,1} = \begin{vmatrix}
        0 & & \\
        -\Tilde{Q}_2 & R_2\\
        & & \ddots
    \end{vmatrix} = 0,
\end{align}
due to the zero on the diagonal. For the minor $M_{2,2}$ we find,
\begin{align}
    M_{2,2} = \begin{vmatrix}
        R_0 & & & & \\
        P_2 & R_2 & & & \\
        & -\tilde{Q}_3 & R_3 & & \\
        & P_4 & -\tilde{Q}_4 & R_4 & \\
        & & & & \ddots
    \end{vmatrix} = R_0 \prod_{i=2}^\infty R_i,
\end{align}
and for minor $M_{2,3}$,
\begin{align}
    M_{2,3} = \begin{vmatrix}
        R_0 & & & & \\
        P_2 & -\Tilde{Q}_2 & & & \\
        & P_3 & R_3 & & \\
        & & -\tilde{Q}_4 & R_4 & \\
        & & & & \ddots
    \end{vmatrix} = -R_0 \Tilde{Q}_2 \prod_{i=3}^\infty R_i,
\end{align}
where we have again made use of Schur's formula shown in Appendix \ref{sec:Mi1}. To see the emergent pattern it is instructive to also calculate $M_{2,4}$, which application of Schur's formula finds as,
\begin{align}
    M_{2,4} = \begin{vmatrix}
        R_0 & & & & & \\
        P_2 & -\Tilde{Q}_2 & R_2 & & & \\
        & P_3 & -\Tilde{Q}_3 & & & \\
        & & P_4 & R_4 & & \\
        & & & -\Tilde{Q}_5 & R_5 & \\
        & & & & & \ddots
    \end{vmatrix} = R_0\begin{vmatrix}
        \tilde{Q}_2 & R_2\\
        P_3 & \tilde{Q}_3
    \end{vmatrix}\cdot \prod_{i=4}^\infty R_i.
\end{align}
If we define $E_i$ as the following determinant,
\begin{align}
    E_i\equiv \begin{vmatrix}
        \Tilde{Q}_2 & R_2 & & & \\
        P_3 & \Tilde{Q}_3 & & & \\
        & & \ddots & & \\
        & & & \Tilde{Q}_{i-1} & R_{i-1} & \\
        & & & P_i & \Tilde{Q}_{i}
    \end{vmatrix},
\end{align}
with $E_0 = 0$ and $E_1 = 1$ we can then express the minors $M_{2,i}$ as,
\begin{align}
    M_{2,i} = E_{i-1} (-1)^i R_0\prod_{j=i}^\infty R_j.
\end{align}
Usage of Cramer's rule from Eq.~\eqref{eq:Cramers} then gives us the elements $\theta_{i,2}$,
\begin{align}
    \theta_{i,2} = \frac{E_{i-1}}{\prod_{j=1}^{i-1}R_j}.
\end{align}
One can also express this in terms of the set of orthogonal polynomials defined in Eq.~\eqref{eq:orthogPolys} to give,
\begin{align}
    \theta_{i,2} = \frac{\phi^{i-1}_{i-2}(0)}{\prod_{j=1}^{i-1}R_j}.
\end{align}

\section{Generating function equations}
\subsection{Asymmetric social choice}\label{sec:appC1}
In the model of asymmetric social choice shown in the main text the generating function PDE corresponding to Eq.~\eqref{eq:ASCme} is given by,
\begin{align}\nonumber
    \partial_t G = &\left((N \varepsilon_1 -\mu_1(N+1))z +(\varepsilon_2+\mu_2(N-1))z^{-1} -N\varepsilon_1 \right)G\\
    & +(z\partial_z)\cdot\left\{\left( (\mu_1(N+1)-\varepsilon_1)z + (\varepsilon_2+\mu_2(N-1))z^{-1} -(\varepsilon_2-\varepsilon_1+N(\mu_1+\mu_2)) \right)  G\right\}\\\nonumber
    & +(z\partial_z)^2\cdot\left\{\left(\mu_1+\mu_2 -\mu_1z-\mu_2 z^{-1} \right) G \right\},
\end{align}
where we have dropped the dependence of $G$ on $z$ and $t$ for brevity, and which was previously studied in \cite{holehouse2022exact}. Using separation of variables, one can show that $G$ has the form $G \sim f(z)e^{-\lambda t}$, which converts this PDE into a second-order ODE in terms of $f_\lambda(z)$. The eigenspectrum $\lambda$ is determined by imposing the finite nature of the population on each $f_\lambda(z)$, i.e., enforcing that it is a polynomial of order $N$ in $z$.

\subsection{Vacillating voters}\label{sec:appC2}
The generating function PDE corresponding to the vacillating voter model is given by,
\begin{equation}\label{eq:voter_ODE}
\begin{split}
\frac{\partial_t G}{N-1}=&\frac{(1-p_d)}{N-1}  z^2(z+1)(z-1) \partial_z^3G\\
&+ \frac{(1-p_d)}{N-1} z(z-1)(2+4z-3Nz)\partial_z^2 G\\
&- (z-1)((N-1)(z+1)p_d+(1-p_d)(N+2z(1-N))) \partial_z G\\
&+ N(z-1)p_dG,
\end{split}
\end{equation}
where we notice the appearance of a third-order term in the generating function ODE whose origin is in the third-order nature of the dynamics of the vacillating voters. Again, one can use separation of variables to convert this PDE into an ODE in terms of $f_\lambda(z)$ and an as yet unidentified spectral parameter $\lambda$,
\begin{equation}
\begin{split}
&(1-p_d)  z^2(z+1)(z-1) \partial_z^3f_\lambda(z)\\
&+ (1-p_d) z(z-1)(2+4z-3Nz)\partial_z^2f_\lambda(z)\\
&- (N-1)(z-1)((N-1)(z+1)p_d+(1-p_d)(N+2z(1-N))) \partial_zf_\lambda(z)\\
&+ (N-1)(\lambda +N(z-1)p_d)f_\lambda(z)=0,
\end{split}
\end{equation}
which is a third-order ODE.

\end{document}